# Improving upon NBA point-differential rankings

Samuel Henry

Abstract

For some time, point-differential has been thought to be a better predictor for future NBA success than pure win-loss record. Most ranking and team performance predictions rely largely on point-differential, often with some normalizations built-in. In this work, various capping and weighting functions are proposed to further improve indicator performance. A gradient descent algorithm is also employed to discover the optimized weighting/capping function applied to individual game scores throughout the season.

## I. Introduction

Up to now, perhaps the most widely used advanced statistical indicator of team ranking is point-differential, or some derivative thereof. This has been thought for some time to be a better predictor for future success than pure win-loss record. Over the years there have been attempts to improve the indicator. Perhaps the most well-known is the Pythagorean Winning percentage from Bill James (developed originally for baseball), applied to basketball [1]. It should be noted, that there are several others available on the web, such as [2, 3].

This work will propose a new methodology for improving this indicator. As opposed to simply averaging each games' outcome (as is done to create point-differential), each function is passed through a weighting function before averaging. To be effective, this function should be non-linear, as a linear weighting function would not change the overall rankings.

First, a hard cap is used to cap each games' differential score before averaging to see if an improved indicator can be used. Second, a gradient descent algorithm is used to find the optimal weight that could be applied to every integer point-margin of victory/defeat. Lastly, specialized functions, commonly used in machine learning, such as the Hyperbolic Tangent function, are used to further improve the indicator. The Pythagorean Winning percentage model applied to NBA will be referenced as a benchmark indicator.

## II. Hard cap solution

The core metric to be analyzed and benchmarked in this paper is the correlation coefficient between *some* indicator, based on data in the *first* half of a given team-season, and the win-loss record in the second half of the team-season. The later part being our proxy for the team performance that we want to predict. Only the regular season is considered here, to include as much data as possible while not biasing or overfitting the results.

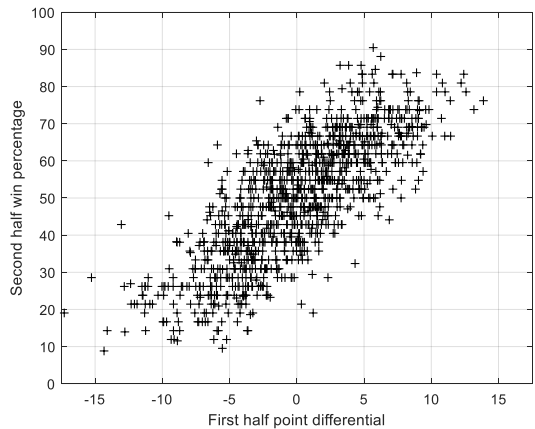

**Figure 1: Correlation between point-differential in first half of season to win-loss record in second half of season**

For example, of what is being discussed, Figure 1 above shows a correlation between point-differential for the first half (*x*-axis), and win-loss record in second half of season (*y*-axis). Data from 1970 and 2014 was used, including a total of 1150 team-seasons, using data from basketball-reference.com.

Then, a simple hard cap is applied to the point-margin of each game before averaging (point-margin is just the differential score for each game; for example, +4 for a 4-point victory). The resulting summation (after dividing by number of games) will be the indicator of future success. For example, if the cap is set at 20 points, and if a team wins by 45 points, there is no additional benefit compared to a 20-point victory. Let's call this Capped Point-differential, $CPD$. The $CPD$ from the first half is used to predict $2^{nd}$ win-loss performance. The question is what cap should be used? The plot below shows the correlation coefficient between $CPD$, and $2^{nd}$ half win-loss record, plotted over cap value.

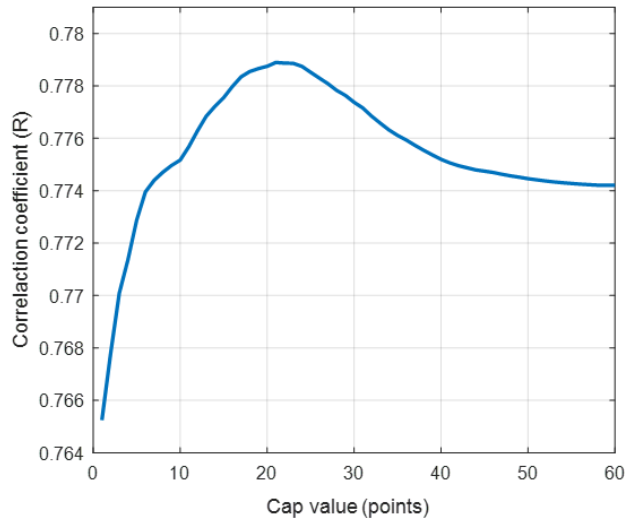

**Figure 2:** *y*-axis is the correlation coefficient between 1st half point differential and 2nd half win-loss record; *x*-axis is the single game cap value used in CPD calculation

Figure 2 is fairly descriptive. Take the left most point for example (76.5% correlation). This represents the case when you cap each point-margin with a +1, or -1. This is the same correlation as using just win-loss record to predict 2nd half win-loss record. Keep in mind using +1 (win) & -1 (loss) vs +1 (win) & 0 (loss) is just a difference in scale factor, which has no effect on the correlation. The far-right portion of the curve converges on using no cap at all, which is the same as using the full point-differential to predict future performance. One can see from the plot that the far-right point is considerably lower than the "peak" correlation.

So, you're giving away a large chunk of the benefit achieved by using point-differential vs simply using win-loss as the indicator. Just by introducing a cap of about 21 points on each game, we can increase the correlation to 77.9%. While this in itself is an interesting finding, let's see if we can do even better.

### III. Weighting every point-margin

The holy grail of this type of indicator would be a weighting function (or perhaps better described as a customized lookup table) that is optimized around each potential point-margin. For example, let's say the $i^{th}$ game of a season a given team wins by 2 points. This game would get a small positive weight. The next game the same team loses by 30. That would be a large negative weight. But how large? That is the essential question to be considered. First, let's define an indicator that sums a weighting value,

$w$, evaluated at every point-margin for the first half of each team-season. The weighted point-differential, $WPD$, is denoted as:

$$WPD = \frac{2}{N} \sum_{i=1}^{N/2} w[pm_i] \qquad (1)$$

where $pm_i$ is the point-margin for the $i^{th}$ game up until game $N/2$ (first half of the season). The traditional point-differential indicator would just be a straight line shown below in Figure 3 on the left side, below. The hard cap solution as discussed in section II would look like something in the right side of Figure 3.

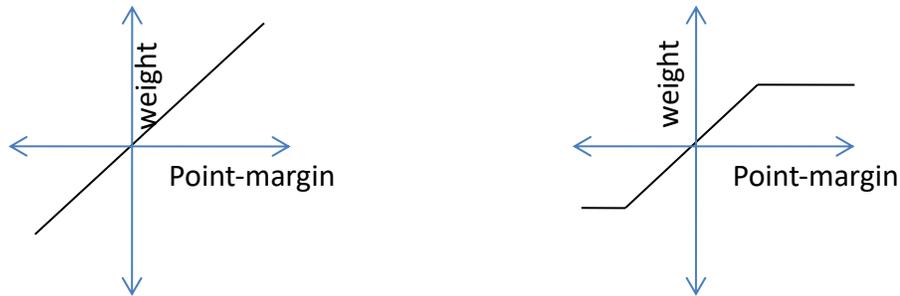

**Figure 3: (left) Linear weighting function results in traditional point-differential-based indicator of future performance. (right) Weighting function for capped point-differential**

The question becomes what does the ideal line (or curve) look like? The most direct way to solve this problem is to bucket each game into counts of point-margins, similarly to a histogram. This is denoted as matrix, $X$. Each of the 1150 team-seasons would have several counts of each possible margin of victory (or defeat). These are then be used as an input to a classic multiple regression. The well-known formula for multiple regression is as follows [4]

$$Y = XW + \varepsilon \qquad (2)$$

where $Y$ is a vector of all win-loss record for the 2nd half of each team-season, so 1150 in length. $W$ is a vector of weights that is ultimately of interest to us. It is 81 units long, with game counts corresponding to -40 to +40 point-margins (wins/losses beyond this were not considered). Here, $X$ is of size 1150 x 81. That is, 1150 team-seasons with 81 different counts of individual point-margins. For example, if a team had amassed 13 4-point wins during team-season #945, then the $X$ [945, 46] would be equal to "13." A gradient descent algorithm is then used to solve the above equation for the weights, $W$, given $X$ and $Y$ [4]. To prevent (or limit) overfitting, a ridge regression value (lambda) of 1 is used.

Figure 4 (left) shows the weights for each point-margin of victory/defeat that results in the maximum correlation to second half win-loss record, namely 80.5%. One can see a relatively linear region in middle, and on the ends (after around ±20 points), the weights become quite noisy. Now one must be careful when interpreting these, as there is likely some overfitting in this function, despite the ridge regression, because there just aren't that many 37-point victories (for example) in the NBA's history.

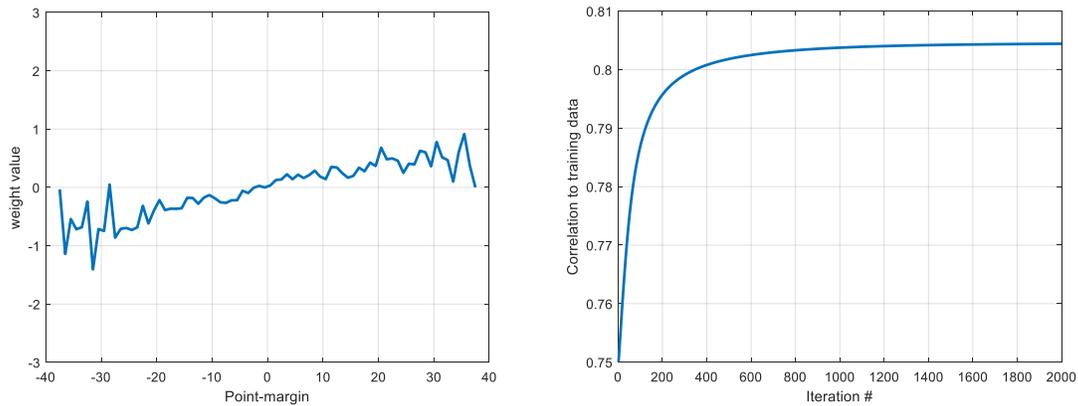

**Figure 4: (left) Plot of optimal weights for each Point-margin after performing gradient. A near-linear region is obvious out until about 15-20 points. (right) Correlation coefficient to training data vs gradient descent iteration #**

But while the resulting correlation coefficient wouldn't completely be a fair (unbiased) indicator of future performance, something can still be learned from the weights themselves. They represent the absolute best possible indicator correlating to win-loss record (trained using past game differential data).

## IV. Soft cap solution

Due to the likely overfitting, this would probably not be an ideal indicator for future performance. But we can use the weights to inspire a new function that would be an improvement over the capped point-differential discussed in section II. Based on Figure 4, it seems clearly there is a "softening" of this linear dependence around points 15-20. At this point the data becomes much noisier, and generally starts to flatten out. But it is reasonable to assume that each additional point of victory has some positive correlation to future performance (likewise for negative). Like in the NBA collective bargaining agreement, a "soft cap" may emerge as the best solution.

Three analytic functions, one of which is heavily used in machine learning a non-linear activation function for artificial neural networks, seem to be a very good fit. Namely, they are the Hyperbolic Tangent function, Error function, and a hand-tuned exponential function. Each function is bounded by 1 and -1 on each side. Each function is essentially linear in the middle, and near horizontal on the extreme positive and negative ends. All three functions are shown in Figure 5 (left). The functions essentially become $w$, used in equation (2).

$$w_{tanh} = \tanh\left(\frac{pm_i}{D}\right) \quad (3)$$

$$w_{erf} = \frac{1}{\sqrt{\pi}} \int_{-pm_i/D}^{pm_i/D} [e^{-t^2}] dt \quad (4)$$

$$w_{exp} = sign[x] \times \left[1 - e^{-|pm_i/D|}\right] \quad (5)$$

The variable, $pm_i$, again is the point-margin for the $i^{th}$ game within any given team-season. The variable, $D$, is a scaling term, which is to be determined. Figure 5 (right) plots correlation coefficient (1st half indicator vs 2nd half win-loss), vs $D$ for all three analytic functions.

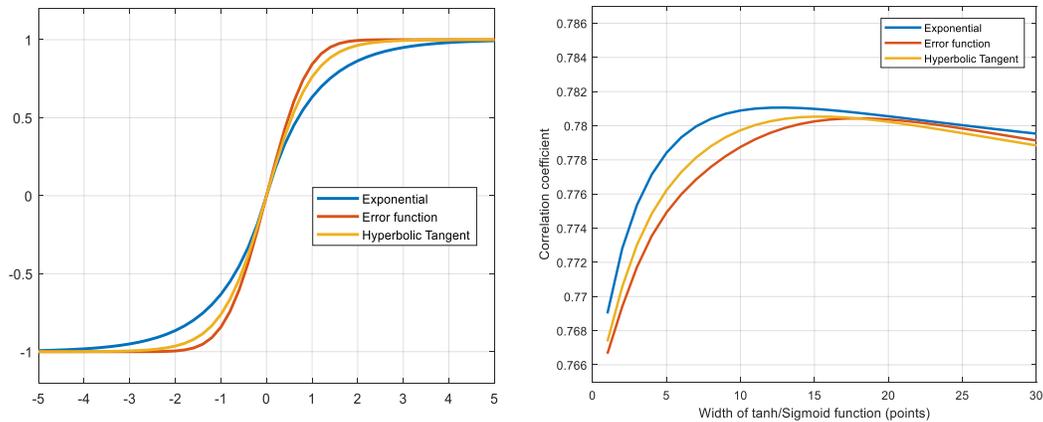

**Figure 5: (left) Exponential, error function, and hyperbolic tangent function plotted with scaling factor (D) equal to unity. (right) Plot of correlation coefficient using soft cap method for all weighting functions**

Similar to the Figure 2, the left most point of all curves in Figure 5 (right) would converge to simply using win-loss record (i.e., when an extremely small width is used, it is like capping each game with a +1, or -1). The right side of all curves would eventually converge on using traditional unweighted point-differential, as this is essentially just using the linear region of the analytic function. However, the plot is zoomed in to show just the peak correlation for all three functions. We can see that the best fit appears to be hand-tuned exponential function with a scaling factor of about 12. The peak correlation for this case lies just above 78.1%.

## V. Conclusion

Table 1 below summarizes all the correlations results so far. For each indicator type that is parameterized (scaling factor or cap, for example), the best fit was chosen. The winner of all the methods tested was the hand-tuned exponential "soft cap" function, with 78.1% correlation.

| Indicator type | Correlation to 2$^{nd}$ half win-loss |
| --- | --- |
| Win-loss | 76.5% |
| Point-differential | 77.4% |
| Capped point-differential | 77.9% |
| Hyperbolic Tangent | 78.1% |
| Error function | 78.0% |
| Exponential | 78.1% |
| Pythagorean Winning Percentage | 77.1% |

Table 1: Comparison of indicators and their respective correlation to 2$^{nd}$ half win-loss record

While these are not monumental differences, it is important to note that the improvement a simple win-loss indicator to point-differential is almost achieved *again* by using the hand-tuned Exponential function. As we continue to get more data over the years (and of course more refined techniques), hopefully there should be improvements on this indicator as well.

What is particularly interesting as we move forward is load management techniques used by teams seem to be gaining steam as more as known about the wear and tear an 82-game season has on the human body. Will resting star players lead to more imbalanced games and "garbage time," and perhaps even less correlation to large victories?

Appendix A: Pythagorean Winning Percentage comparison

In Table 1, the Pythagorean Winning Percentage was used as a baseline metric to compare to other indicators of future performance. The exponent used was 2.4. This number was imperially determined based on the same data set used in the rest of the work, from 1970 and 2014, including a total of 1150 team-seasons, from basketball-reference.com. See figure below for correlation (R) plotted as a function of the exponent, *exp*. A maximum is visible at around 2.4.

$$PYTHAGOREAN\ WINNING\ \% = \frac{[PTS\ SCORED]^{exp}}{[PTS\ ALLOWED]^{exp} + [PTS\ SCORED]^{exp}}$$

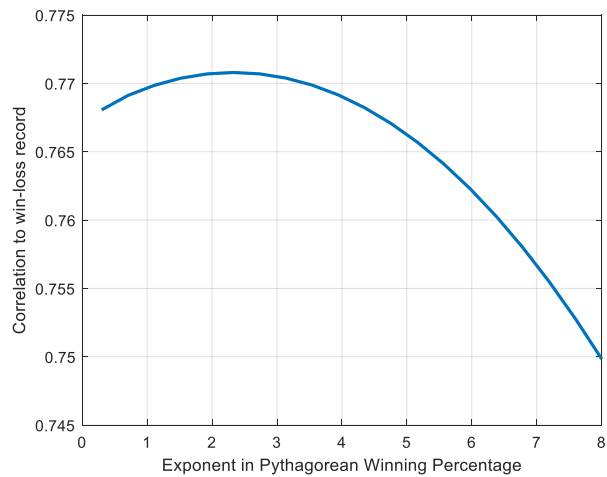

**Figure 6: Plot of correlation coefficient using Pythagorean winning percentage versus exponent, *exp*, used in calculation**

# Citations